\documentclass[a4paper,12pt,onecolumn]{article}
\usepackage{graphicx} 
\usepackage{booktabs}
\usepackage{authblk}
\usepackage{graphicx}
\usepackage{physics}
\usepackage{url}
\usepackage[breaklinks]{hyperref}
\usepackage{geometry}
\usepackage{ragged2e}
\usepackage[T1]{fontenc}
\usepackage{lipsum}
\usepackage{authblk}
\usepackage{titlesec}
\usepackage{caption}
\usepackage{units}
\usepackage{ifthen}
\usepackage{blindtext}
\usepackage{fancyhdr}
\usepackage{soul}
\sethlcolor{yellow}
\usepackage[backend=biber,style=nature,sorting=none,maxnames=4]{biblatex}
\usepackage[shortcuts]{extdash}
\appto\bibfont{\setlength{\emergencystretch}{.5em}}
\titleformat{\paragraph}[runin]
  {\itshape}
  {}
  {0pt}
  {}

\date{}

\usepackage{sectsty}
\subsectionfont{\sffamily}

\usepackage[normalem]{ulem}

\usepackage[dvipsnames,table,xcdraw]{xcolor}
\definecolor{SciAdv}{rgb}{0.678, 0.051, 0.078}
\sectionfont{\color{SciAdv}}
\usepackage{hyperref}
\hypersetup{
    colorlinks=true,
    citecolor=blue,
    linkcolor=blue,
    filecolor=magenta,      
    urlcolor=blue
}

\newcommand{\iu}{\mathrm{i}}
\newcommand{\eu}{\mathrm{e}}

\usepackage{csquotes}

\usepackage{amsmath}
\usepackage{enumitem}

\begin{document}

\title{From Hollow Waveguides to Mie Voids: \\Low-Index Light Confinement}

\renewcommand\Authfont{\normalsize\raggedright}
\renewcommand\Affilfont{\small\itshape\raggedright}
\setlength{\affilsep}{1em}
\author[1]{D. Ludescher}
\author[2]{I. Toftul}
\author[1]{H. Giessen}
\author[2,*]{Y. Kivshar}
\author[1,*]{M. Hentschel}
\affil[1]{4th Physics Institute and Research Center SCoPE, University of Stuttgart, 70569 Stuttgart, Germany}
\affil[2]{Research School of Physics, Australian National University, 2600 Canberra ACT, Australia}
\affil[*]{Corresponding authors: Yuri Kivshar (yuri.kivshar@anu.edu.au) and Mario Hentschel (m.hentschel@pi4.uni-stuttgart.de)}

\maketitle

\subsection*{Abstract}
The confinement of light within optically thin media challenges the conventional view that optical localization requires high-index dielectric materials and total internal reflection. Over the past decades, a broad range of photonic phenomena has demonstrated that electromagnetic energy can also be guided and localized inside low-index regions or nanoscale void cavities. What first appears to be a curiosity is, in fact, highly relevant: The interaction between light and matter lies at the heart of much in optics and photonics. It enables sensing, nonlinear optical phenomena, and optical manipulation, yet it also inevitably introduces absorption, dissipation, and loss. Pushing localized, enhanced, or guided energy into a lower-index space thus offers entirely new possibilities for light-matter interaction and dispersion control. In this review, we discuss some of these counterintuitive structures, starting with light guidance within optically thin media, such as hollow-core photonic crystal fibers or antiresonant reflecting optical waveguides. We then discuss the optics of bubbles, that is, low-index inclusions in a higher-index surrounding, and ultimately arrive at the so-called Mie void resonators, optical resonators that support localized electromagnetic modes confined within a region of lower refractive index. All these systems demonstrate that there are (resonant) optics beyond simple total internal reflection. Absence can itself become a functional photonic degree of freedom, opening opportunities across sensing, quantum nanophotonics, metasurfaces, and programmable optical matter.
\newpage
\section*{Introduction}
The ability to guide and confine optical energy within low-index regions has attracted growing interest as it offers a fundamentally different way of controlling electromagnetic fields. Unlike a familiar Fabry-P\'erot resonator geometry where light is bounded by mirrors, in most conventional optical systems light resides inside optically dense materials, where confinement emerges through refractive-index contrast and total internal reflection. Confining unbounded light in air appears counterintuitive. Yet this inversion gives rise to optical behaviors and functionalities that are not readily accessible in conventional dielectric structures.\\
Many of these opportunities arise from the direct spatial accessibility of the confined optical fields. If optical fields can be localized directly within air-filled or otherwise accessible regions, surrounding matter can naturally overlap with the electromagnetic hot spots. Gases, liquids, nanoparticles, molecules, and quantum emitters can then interact directly with strongly localized optical fields without requiring penetration into dense dielectric materials. At the same time, reducing the overlap between light and absorbing matter can suppress losses, nonlinear absorption, heating, and material damage. Systems capable of guiding or localizing light in low-index environments therefore promise unusual combinations of many properties such as accessibility, robustness, and flexible optical control.\\
Beyond these considerations, the confinement of light in optically thin media is also intrinsically interesting from a fundamental perspective. Electromagnetic localization has traditionally been associated with \enquote{dielectric accumulation}, where optical energy remains trapped inside regions of high refractive index. Structures that confine light in air challenge this intuition by demonstrating that localization can also emerge from interference, resonance formation, inhibited coupling, and inverse dielectric geometries. We formalize this intuition as a variational principle and show where it fails. The ability to localize light within carefully engineered \enquote{voids} therefore reflects a broader expansion in how photonics approaches the control of electromagnetic fields.\\
Here we review a range of approaches for guiding and localizing light within optically thin media. We begin with waveguiding concepts based on interference, inhibited coupling, and structured electromagnetic environments before turning to inverse dielectric scattering systems and resonators. Particular emphasis is placed on \emph{dielectric Mie voids} as a platform for studying inverse dielectric resonances and their applications in sensing, light-matter interaction, and nanophotonics. Finally, we discuss the broader implications and future opportunities arising from optical confinement in low-index environments.
\section{Beyond total internal reflection}
\label{sec:beyond}
\begin{figure}[t!]
\centering
\includegraphics[width=\linewidth]{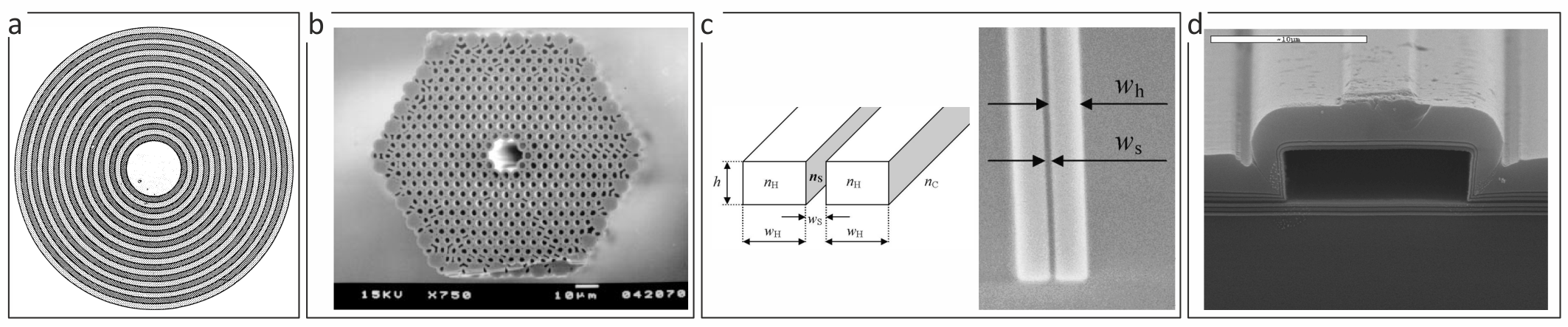}
\caption{\textbf{Guiding light in optically thin media.} \textbf{a}, Schematic of a Bragg fiber consisting of an air-filled core surrounded by concentric dielectric layers that confine light through Bragg reflection. \textbf{b}, Scanning electron micrograph of a hollow-core photonic crystal fiber, where guidance is achieved through a photonic bandgap. \textbf{c}, Schematic diagram and scanning electron micrograph of the slot-waveguide geometry illustrating field concentration within a nanoscale low-index gap between two high-index regions. \textbf{d}, Scanning electron micrograph of an antiresonant reflecting optical waveguide (ARROW). a, Reprinted with permission.\supercite{Yeh1978} Copyright 1978 Optica Publishing Group. b, Adapted with permission.\supercite{Cregan1999} Copyright 1999 American Association for the Advancement of Science (AAAS). c, Adapted with permission.\supercite{Xu2004} Copyright 2004 Optica Publishing Group. d, Reprinted with permission.\supercite{Yin2004_2} Copyright 2004 AIP Publishing LLC.}
\label{fig:Fig1}
\end{figure}

One of the earliest routes toward low-index optical confinement emerged from the engineering of the surrounding environment. It can be classified as Fabry-P\'erot-type localization. Here, a \enquote{mirror} is meant in a broad sense: any surrounding structure that reflects light back into the low-index region, from a periodic Bragg stack to a single interface.\\
Yeh and Yariv introduced Bragg-guided waveguides as one of the earliest demonstrations that optical confinement can emerge from distributed interference.\supercite{Yeh1976,Yeh1978,Yeh1988,Fink1998,Temelkuran2002} In these systems, periodic dielectric environments suppress radiative propagation even when the guided region itself possesses a comparatively low refractive index. The feedback is provided by a periodic stack of materials with different dielectric constants (Bragg stack). Reflections at carefully designed interfaces induce constructive interference, leading to local confinement that no longer arises solely from evanescent exclusion at a single interface, but instead emerges from the structure of the surrounding medium. The optical environment thereby becomes an active component in determining whether radiative channels remain accessible at all.\\
For a cylindrical geometry, this conceptual idea is pushed forward in so-called Bragg fibers, illustrated in Fig.~\ref{fig:Fig1}a.\supercite{Yeh1978} Light can remain localized inside the air-filled core provided the surrounding electromagnetic environment sufficiently suppresses leakage into radiative states, in this case by the periodic modulation of the dielectric constants of the radially symmetric \enquote{fiber shell}. Although technologically demanding, it is an interesting topic which is still pushed in experiment and modeling.\supercite{Brechet2000,Engeness2003,Johnson2001}
Russell and co-workers later extended this concept through photonic crystal fibers, where wavelength-scale structuring of the cladding environment enables hollow-core guidance within air-filled regions,\supercite{Cregan1999,Russell2003,Russell2006} as illustrated by the scanning electron microscope image displayed in Fig.~\ref{fig:Fig1}b. Here, the cladding is no longer a passive dielectric background, but possesses a photonic band structure capable of suppressing radiative escape from the core. In hollow-core photonic crystal fibers, it is not the (air-)core itself that is intrinsically guiding, but instead guidance arises because the surrounding dielectric environment restricts the optical states available for radiative propagation.\\
At the same time, it was found that the low-index confinement mechanisms for photonic crystal fibers often resist clean separation into distinct physical categories. Depending on geometry and spectral regime, confinement could be interpreted through photonic bandgaps, modified total internal reflection, distributed interference, or inhibited coupling between core and cladding states.\supercite{Cregan1999,Russell2003,Russell2006,Benabid2002,Couny2006,Poletti2014,Ding2014,Debord2017,Joannopoulos2008}\\
Related concepts also emerged in inverse opal photonic crystals composed of periodic air cavities embedded within high-index dielectric materials.\supercite{Wijnhoven1998} In these systems, the optical functionality arises primarily from collective photonic band-structure effects generated by periodically arranged low-index inclusions. Although inverse opals are not generally discussed in terms of localized resonant confinement inside accessible voids, they further reinforce the broader idea that carefully designed void structures can strongly reshape electromagnetic propagation and optical state formation.\\
To understand where light localizes, a variational principle provides a useful starting point. We consider linear, isotropic, lossless, and time-invariant media. They are non-dispersive and non-magnetic, with relative permeability $\mu=1$, and their relative permittivity \(\varepsilon=\varepsilon(\mathbf{r})\) is real and positive. 
All fields vary in time as $\exp(-\iu\omega t)$, and $\mathbf{E}(\mathbf{r})$ is the complex amplitude of the electric field. A bound mode has a real frequency $\omega$ and radiates no energy to infinity. The lowest bound mode obeys\supercite{Joannopoulos2008}
\begin{equation}
	\label{eq:variational}
	\frac{\omega^2}{c^2}=\min_{\nabla\cdot(\varepsilon\mathbf{E})=0}\frac{\int|\nabla\times\mathbf{E}|^2\,\dd V}{\int\varepsilon|\mathbf{E}|^2\,\dd V},
\end{equation}
where $c$ is the speed of light in vacuum. The integrals run over all space. For a waveguide, they run over the cross section at a fixed propagation constant. 
One way to minimize Eq.~\eqref{eq:variational}  is to concentrate the field in the regions of the high values of \( \varepsilon \) (maximizes the denominator, which is the electric energy, up to the factor $\varepsilon_0/4$) and to have as few spatial oscillations as possible (minimizes the numerator).
Waveguides that use total internal reflection follow this route. However, it is not the only way of minimizing Eq.~\eqref{eq:variational}. 
The boundary condition keeps the normal displacement \( D_{\perp} = \varepsilon_{0}\varepsilon(\mathbf{r}) E_{\perp} \) continuous. A field normal to an interface is therefore \( \varepsilon_{\text{ex}} / \varepsilon_{\text{in}} \) times stronger on the inner side, \( E_{\perp}^{\text{in}} = (\varepsilon_{\text{ex}}/\varepsilon_{\text{in}})\, E_{\perp}^{\text{ex}} \). Consider now a low-index gap, \( \varepsilon_{\text{in}} < \varepsilon_{\text{ex}} \), between high-index parts, and let the gap be narrower than the decay length of the field. If the field fell inside the gap, it would vary strongly in space and raise the numerator of Eq.~\eqref{eq:variational}. The field therefore keeps its enhanced value across the gap, and is maximal in the hollow region. This is the main operational principle of slot waveguides.\supercite{Almeida2004,Xu2004}\\
Almeida and co-workers demonstrated that strong electric-field enhancement can occur inside nanoscale low-index gaps separating high-index dielectric regions,\supercite{Almeida2004,Xu2004} illustrated in Fig.~\ref{fig:Fig1}c. Unlike the Bragg guidance or antiresonant reflection discussed before, this localization is a direct consequence of Maxwell's boundary conditions and the continuity of the displacement field across large refractive-index discontinuities. A \unit[50]{nm} silica gap between silicon rails carries \unit[30]{$\%$} of the guided power, and the mode is bound.\supercite{Almeida2004} What makes these structures fundamentally important is the insight that in such slot and void waveguides, low-index regions themselves can hold the highest electromagnetic energy density.\\
Duguay and co-workers introduced so-called antiresonant reflecting optical waveguides (ARROW) as structures in which confinement emerges through phase-sensitive suppression of leakage into surrounding dielectric channels,\supercite{Duguay1986,Yin2004_2,Yin2004} as depicted in Fig.~\ref{fig:Fig1}d. The surrounding interfaces behave as antiresonant reflectors whose optical response minimizes coupling between the guided mode and the exterior environment. Importantly, these systems remain intrinsically leaky in a strict electromagnetic sense, emphasizing that practical low-index confinement frequently relies not on perfectly bound states, but on sufficiently suppressing radiative escape over technologically relevant propagation
 lengths.\\
As antiresonant and hollow-core systems evolved further, researchers increasingly shifted the language of confinement away from complete photonic bandgaps toward modal accessibility and inhibited coupling. Benabid and co-workers, together with later developments on Kagome, negative-curvature, and antiresonant hollow-core fibers, demonstrated that confinement can persist even when numerous radiative states remain available within the surrounding dielectric environment, provided the overlap between guided and radiative modes remains sufficiently weak.\supercite{Benabid2002,Couny2006,Poletti2014,Ding2014,Debord2017,Joannopoulos2008} Low-index guidance therefore became a problem not simply of preventing propagation in the cladding, but of engineering how efficiently electromagnetic states can communicate with one another.
\section{From bubbles to resonant voids}
\label{sec:bubbles}
\begin{figure}[t!]
\centering
\includegraphics[width=\linewidth]{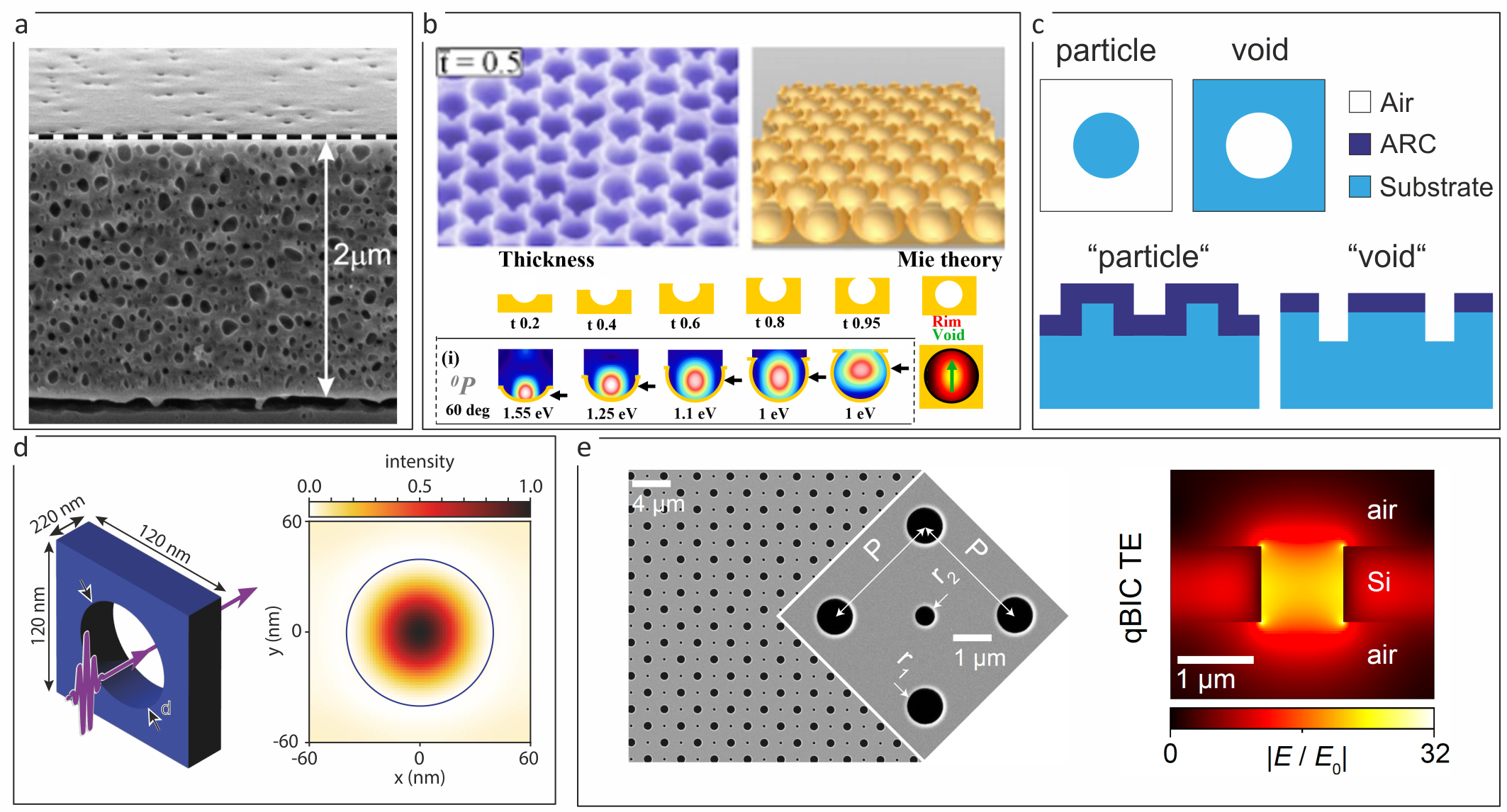}
\caption{\textbf{Optical systems based on low-index inclusions and voids.} \textbf{a}, Scanning electron micrograph of porous silicon containing a high density of air voids embedded within a silicon matrix. \textbf{b}, Thickness variations in metallic nanoscale voids are investigated, with particular emphasis on the coupling between void plasmons and rim dipoles. \textbf{c}, Comparison of dielectric particle and dielectric void resonators proposed for light trapping in thin-film solar cells (ARC: anti-reflection coating). \textbf{d}, Schematic of an extreme-ultraviolet dielectric meta-atom consisting of a nanohole etched into a silicon membrane together with the simulated electric-field distribution at resonance. These examples demonstrate how low-index inclusions can modify optical scattering, absorption, and resonant behavior. \textbf{e}, Scanning electron micrographs of a quasi-bound-state-in-the-continuum (qBIC) metasurface and its unit cell together with the simulated electric-field distribution of the supported mode. These examples illustrate different approaches for confining and guiding light within low-index regions. a, Adapted with permission.\supercite{Seel2004} Copyright 2004 Elsevier B.V. b, Adapted with permission.\supercite{Cole2007} Copyright 2007 American Chemical Society. c, Adapted with permission.\supercite{Mann2011} Copyright 2011 Optica Publishing Group. d, Adapted with permission.\supercite{Ossiander2023} Copyright 2023 American Association for the Advancement of Science (AAAS). e, Adapted with permission.\supercite{Adi2024} Copyright 2024 Springer Nature Limited.}
\label{fig:Fig2}
\end{figure}

Moving away from guided modes, we now consider compact resonators, hence a spherical symmetry of the mirror. 
The quality factor of the radial mode of a spherical void (radius $R$, index $n_\mathrm{in}$) in a host with refractive index $n_\mathrm{ex} > n_\mathrm{in}$ can be estimated within a Fabry-P\'erot picture. Lacking total internal reflection, the mode is confined only by normal-incidence Fresnel reflection at the boundary, losing the fraction $1-|r|^2 = 4n_\mathrm{in}n_\mathrm{ex}/(n_\mathrm{in}+n_\mathrm{ex})^2$ of its energy per round trip time $t_\mathrm{r.t.} = 2n_\mathrm{in}R/c$. Then $Q_{\text{void}} = \omega t_\mathrm{r.t.}/(1-|r|^2)$ for $n_\mathrm{ex} \gg n_\mathrm{in}$ gives
\begin{equation}
	Q_{\text{void}} \approx  \pi\,\frac{R/\lambda}{n_\mathrm{in}/n_\mathrm{ex}},
	\label{eq:Q_void}
\end{equation}
with $\lambda = \lambda_0/n_\mathrm{in}$. In the ray picture, $Q_{\text{void}}$ grows linearly with both the void size and the refractive index contrast. We note that such resonators are inherently leaky and hence the variational principle in Eq.~\eqref{eq:variational} is not applicable here anymore.\\
Among the simplest examples of an inverse dielectric system is an air bubble suspended in water, where a low-index region is completely surrounded by a material of higher refractive index. Variations of this geometry appear in many other contexts, including porous semiconductors, solar-cell textures, metallic nanocavities, and inverse dielectric resonators. Having a common geometry, they provide useful reference points for understanding how voids and cavities can acquire structured optical responses.\\
The optical properties of air bubbles have been studied extensively, primarily in the context of scattering and metrology. Typical bubbles are much larger than visible wavelengths, often ranging from tens of micrometers to millimeters, and the refractive-index contrast between air and water is comparatively small.\supercite{Rodrigues2003} They are therefore not expected to support nanoscale resonant confinement. Patitsas and co-workers analyzed the scattering of spherical voids and revealed that angular extrema in the Mie scattering curve can be used to determine bubble sizes.\supercite{Patitsas1976} Marston and collaborators later used Mie theory and ray-optical descriptions to study critical-angle scattering, diffraction effects, angular oscillations, and morphology-dependent features in bubbles.\supercite{Marston1979,Kingsybury1981,Marston1982} These studies used inverse dielectric Mie scattering mainly to describe angular scattering patterns and metrology signals rather than localized cavity modes. Related work further established light scattering as a method for characterizing air bubbles in water.\supercite{Patitsas1976,Hansen1985}\\
A separate body of work considered voids and nanovoids in semiconductor materials. In amorphous and porous silicon, low-index inclusions were studied because they modify effective optical constants, scattering, absorption, and light transport.\supercite{Vanecek1991} Seel and Brendel investigated porous silicon layers and the influence of voids in crystalline silicon, depicted in Fig.~\ref{fig:Fig2}a and presented that voids can strongly affect internal scattering and absorption in thin semiconductor structures.\supercite{Seel2004} Related studies on nanovoid-containing silicon films further underpinned that ensembles of voids can enhance absorption through multiple scattering and light trapping.\supercite{Vanecek1991,Banerjee2005_1} More recently, inverse dielectric scattering in absorbing host media has been revisited using generalized Mie theory,\supercite{Zhang2023} presenting that host absorption can substantially modify extinction and attenuation by embedded voids. In these systems, voids are not treated primarily as isolated resonant cavities, but as structural elements that modify optical transport in absorbing  media.\\
A particularly relevant parallel development occurred in plasmonics, where metallic nanovoids were investigated as complementary counterparts to metallic nanoparticles. Coyle and co-workers demonstrated that near-spherical nanocavities in a gold support layer exhibit localized surface-plasmon resonances whose energies can be tuned from the ultraviolet to the near-infrared region by varying cavity diameter and height.\supercite{Coyle2001} Importantly, the measured resonance energies were found to agree with the Mie resonances of a spherical void in a metallic host. This concept was subsequently developed in greater detail by Kelf and co-workers, who distinguished localized \enquote{Mie plasmons} of truncated spherical nanovoids from delocalized Bragg plasmon modes of the patterned metal surface.\supercite{Kelf2006} Their analysis reveals how localized dipolar and higher-order void modes evolve as the cavity geometry changes from a shallow dish to an almost fully enclosed spherical void.\\
Cole and co-workers later provided a more complete description of these metallic void resonances, referring to them as plasmonic \enquote{anti-nanoparticles}.\supercite{Cole2007} For a fully enclosed spherical cavity, the resonant solutions can again be obtained from Mie theory, as can be seen from the rightmost field distribution in Fig.~\ref{fig:Fig2}b (bottom). The fully closed metallic shell supports a well-defined cavity mode, fundamentally originating from the mirror-cavity formed by the gold shell. As soon as the void opens, as will be the case in the experiment, the metal surface introduces additional rim-plasmon modes. These can hybridize with the intrinsic void modes to form bonding and antibonding states with strongly enhanced local fields, increasingly deviating from the simple Mie-like modes for the perfect spherical void with increasing truncation. This transition is illustrated by the additional field distributions in Fig.~\ref{fig:Fig2}b. Related work extended these ideas to coupled nanovoids, where hybridization between individual void plasmons creates plasmonic molecular states,\supercite{Romero2008} as well as to diffraction enhancement, near-field imaging, surface-enhanced Raman scattering (SERS), and hybrid material systems.\supercite{Teperik2006,Lacharmoise2008,Cole2009,Tognalli2012,Zhu2013} Romero and co-workers, for example, demonstrated that overlapping buried voids in gold can form coupled plasmonic states whose resonance energies are tuned through the void overlap. Here, coupling is mediated by the overlap of the modal regions in air, providing a complementary analogue of coupled nanoparticle plasmonics. Zhu and co-workers have also used the enhanced fields to boost SERS signals from graphene. In this experiment, the graphene layer is placed on top of the void array and thus overlaps with the accessible field enhancement inside the voids.\supercite{Zhu2013}\\
Due to experimental challenges, the implementation has generally resulted in truncated voids in a metallic layer, deviating from the ideal geometry. Recently, however, fully enclosed three-dimensional aluminum nanovoids could be realized in chemically synthesized nanoparticles. In these systems, the mode formation is also partially governed by plasmonic resonances and mode hybridization, leading to significant field localization inside the metal. Interestingly, these localized plasmon resonances can nevertheless extend into the extreme-ultraviolet regime, further demonstrating that the concept of the void as a resonant optical object extends well beyond noble-metal surface cavities.\supercite{Zhu2017}\\
A more direct connection to resonant dielectric nanophotonics was made by Mann and co-workers, who explicitly compared dielectric particles and dielectric voids as complementary resonator geometries for thin-film solar-cell textures.\supercite{Mann2011} In this case, as illustrated in Fig.~\ref{fig:Fig2}c, the relevant dimensions on the order of the optical wavelength allow for resonant scattering effects to contribute to light trapping. The study confirms that low-index cavities embedded in high-index media can support resonant optical responses and forward scattering effects, analogous in some respects to conventional dielectric particles. One of the most important findings that can be deduced from these results is the realization that a mirror-cavity does not necessarily require metallic mirrors, but can also be achieved due to a purely dielectric contrast. The benefit is not only a reduced complexity in the system, but also the absence of plasmonic modes hybridizing with the Mie-type cavity modes.\\
The existence of localized resonances in inverse dielectric systems is, however, physically distinct from the metallic case. In metallic nanovoids, the modes originate from surface-charge oscillations at the metal-void interface and may hybridize with rim or propagating surface-plasmon modes.  In dielectric systems, no free-electron resonance is available, and no strict Babinet symmetry exists. Mie theory itself imposes no requirement that the inclusion possess a higher refractive index and formally describes both particles and voids, while the actual resonance condition follows from the full electromagnetic boundary-value problem. Babinet-type arguments therefore provide a useful conceptual reference for complementary metallic structures,\supercite{Falcone2004,Zentgraf2007,Cole2007} but only an approximate analogy for dielectric Mie voids.\supercite{Hamidi2025}\\
A related but distinct example appears in extreme-ultraviolet nanophotonics. Ossiander and co-workers used nanoholes in thin silicon membranes as meta-atoms for operation at a wavelength of \unit[50]{nm}.\supercite{Ossiander2023} In this spectral regime, the refractive index of silicon is below that of vacuum, so the vacuum-filled hole becomes the optically denser region, illustrated in Fig.~\ref{fig:Fig2}d. The field localization inside the nanohole therefore corresponds to a conventional Mie-type resonance in the optically denser medium, rather than to inverse dielectric confinement. This is a particularly elegant design that plays on intuition and is once again a demonstration of the many options available to structure and guide light flow on the nanoscale.\\
More recently, metasurfaces supporting quasi-bound states in the continuum (qBICs) have further extended the localization of optical fields inside accessible low-index regions. Free-standing dielectric membrane metasurfaces (see left of Fig.~\ref{fig:Fig2}e) can support high-$Q$ resonances whose electromagnetic hot spots reside predominantly within air cavities rather than inside the surrounding dielectric environment (compare simulated field distribution on the right of Fig.~\ref{fig:Fig2}e).\supercite{Adi2024}  These systems again illustrate that optical localization in low-index media can emerge from carefully engineered radiative coupling and interference rather than from conventional total internal reflection alone.
\section{Fundamentals of Mie voids}
\subsection*{Derivation of Mie resonances}
Mie theory solves the scattering of a plane wave by a homogeneous sphere exactly.\supercite{Mie1908,BohrenHuffman1983} We consider a sphere of radius $R$ and refractive index $n_{\mathrm{in}}$ in a host of index $n_{\mathrm{ex}}$. Both media are homogeneous, isotropic, and nonmagnetic, and the fields vary in time as $\exp(-\iu\omega t)$. Two dimensionless numbers fix the problem. We use the size parameters $x=n_{\mathrm{ex}}k_0R$ and $y=n_{\mathrm{in}}k_0R$, with $k_0=\omega/c$, and their ratio, the relative index $m=n_{\mathrm{in}}/n_{\mathrm{ex}}=y/x$. A particle has $m>\text{1}$, and a void has $m<\text{1}$. The theory is the same for both.\\
The derivation has three steps. First, in each medium, Maxwell's equations reduce to the vector Helmholtz equation $\nabla\times\nabla\times\mathbf{E}=n^2k_0^2\mathbf{E}$, with $n=n_{\mathrm{in}}$ or $n_{\mathrm{ex}}$. Second, the spherical symmetry separates its solutions into spherical multipoles of order $\ell=\text{1,2,}\ldots$, namely vector spherical harmonics.  Its radial dependence is expressed via spherical Bessel functions. Inside, the solution must be regular at $r=\text{0}$, which selects $j_\ell$. Outside, the scattered field must be an outgoing wave, which selects spherical Hankel functions $h_\ell^{(1)}$. A plane wave contains all possible $\ell$ with known weights.\supercite{BohrenHuffman1983} 
We decompose incident, scattered, and internal fields into a series of vector spherical harmonics. After that, we apply boundary conditions on the particle surface, and at infinity for the scattered field.
Eliminating the internal amplitudes,
in terms of the Riccati-Bessel functions $\psi_\ell(z)=zj_\ell(z)$ and $\xi_\ell(z)=zh_\ell^{(1)}(z)$, one can find
\begin{equation}
\label{eq:a}
a_\ell=\frac{m\psi_\ell(y)\psi_\ell'(x)-\psi_\ell(x)\psi_\ell'(y)}{m\psi_\ell(y)\xi_\ell'(x)-\xi_\ell(x)\psi_\ell'(y)}\equiv\frac{N_\ell^{\mathrm{TM}}}{D_\ell^{\mathrm{TM}}}
\end{equation}
and
\begin{equation}
\label{eq:b}
b_\ell=\frac{\psi_\ell(y)\psi_\ell'(x)-m\psi_\ell(x)\psi_\ell'(y)}{\psi_\ell(y)\xi_\ell'(x)-m\xi_\ell(x)\psi_\ell'(y)}\equiv\frac{N_\ell^{\mathrm{TE}}}{D_\ell^{\mathrm{TE}}}.
\end{equation}
Each coefficient is normalized such that $|a_{\ell}|,|b_{\ell}|<\text{1}$.
The denominator follows from the numerator by the replacement $\psi_\ell(x)\to\xi_\ell(x)$, that is, by replacing the incident wave with the outgoing scattered wave.\\
To determine the intrinsic electromagnetic eigenmodes of the sphere, we consider the absence of an incident field. Consequently, the driving terms proportional to the incident amplitude vanish, leaving a homogeneous boundary system. Nontrivial solutions exist only when the corresponding determinant vanishes. These determinants coincide with the denominators $D_\ell^{\mathrm{TM}}$ and $D_\ell^{\mathrm{TE}}$ of the Mie scattering coefficients (Eqs.~\eqref{eq:a} and~\eqref{eq:b}), leading to the resonance conditions
\begin{equation}
\label{eq:resonance}
D_\ell^{\mathrm{TM}}(\tilde{\omega})=0, \qquad
D_\ell^{\mathrm{TE}}(\tilde{\omega})=0.
\end{equation}
For passive, radiating resonators, the resulting eigenfrequencies are generally complex,\linebreak $\tilde{\omega}=\omega_{\mathrm{r}}-\iu\gamma/2$. Here, $\omega_{\mathrm{r}}=\Re(\tilde{\omega})$ denotes the resonance frequency, whereas \linebreak $\gamma=\gamma_{\mathrm{rad}}+\gamma_{\mathrm{abs}}=-2\Im(\tilde{\omega})$ characterizes the energy decay rate, with radiative and absorptive contributions $\gamma_{\mathrm{rad}}$ and $\gamma_{\mathrm{abs}}$. Accordingly, the eigenfrequencies lie in the lower half of the complex frequency plane. The quality factor $Q=\omega_{\mathrm{r}}/\gamma$ of Mie void resonances is determined by the resonance frequency relative to the total energy decay rate. Lower radiative and absorptive losses therefore result in higher quality factors, corresponding to longer resonance lifetimes and narrower spectral linewidths.\\
In the vicinity of a simple resonance, provided that the corresponding numerator is nonzero at the pole and neglecting the regular background term, the scattering coefficient can be approximated by
\begin{equation}
\label{eq:pole}
a_\ell(\omega)\approx\frac{N_\ell^{\mathrm{TM}}(\tilde{\omega})}{\omega-\tilde{\omega}}\left(\left.\frac{\dd D_\ell^{\mathrm{TM}}}{\dd\omega}\right|_{\omega=\tilde{\omega}}\right)^{-1}.
\end{equation}
The complex eigenfrequencies can be determined numerically by applying Newton's method to the denominators in Eq.~\eqref{eq:resonance},
\begin{equation}
\label{eq:newton}
\omega_{j+1}=\omega_j-D(\omega_j)\left(\left.\frac{\dd D}{\dd\omega}\right|_{\omega=\omega_j}\right)^{-1},
\end{equation}
where $D$ represents either $D_\ell^{\mathrm{TM}}$ or $D_\ell^{\mathrm{TE}}$.\\
Thus, zeros of the denominators generally correspond to poles of the scattering coefficients, provided that they are not canceled by simultaneous zeros of the numerators. Conversely, zeros of the numerators indicate suppression of the corresponding scattering multipoles when the denominators remain finite. Importantly, the spectral maxima observed along the real frequency axis do not necessarily coincide with the real parts of the complex eigenfrequencies. In dispersive media, the refractive indices $n_{\mathrm{in}}(\omega)$ and $n_{\mathrm{ex}}(\omega)$ must additionally be analytically continued into the complex frequency plane. The same theoretical framework applies to Mie voids, characterized by an internal refractive index lower than that of the surrounding medium, $n_{\mathrm{in}}<n_{\mathrm{ex}}$ ($m<\text{1}$).\supercite{Hamidi2025}\\
In the limiting case of a lossless host with increasingly large refractive-index contrast,\linebreak $n_{\mathrm{ex}}/n_{\mathrm{in}}\rightarrow\infty$ ($m\rightarrow\text{0}$), while maintaining a fixed internal size parameter $y$, the external size parameter $x$ becomes asymptotically large and the relation
\begin{equation}
\label{eq:limit}
\frac{\xi_\ell'(x)}{\xi_\ell(x)}\rightarrow\iu
\end{equation}
simplifies the resonance conditions~\eqref{eq:resonance} to
\begin{equation}
\label{eq:limit-res}
\begin{split}
\text{TM:}&\quad\psi_\ell'(y)=\text{0},\\
\text{TE:}&\quad\psi_\ell(y)=\text{0}.
\end{split}
\end{equation}
These are the resonance conditions of a cavity with a perfectly conducting wall. In this limit, the void mode becomes a bound mode of a closed cavity. For the lowest multipolar order ($\ell=\text{1}$), the first positive roots are
\begin{equation}
\label{eq:roots}
\begin{split}
y_{\mathrm{TM},\ell=1}&\approx \text{2.7437},\\
y_{\mathrm{TE},\ell=1}&\approx \text{4.4934}.
\end{split}
\end{equation}
These two values represent the dimensionless eigenvalues of the lowest-order electromagnetic multipolar resonances of a sphere. They can be used to approximate the corresponding resonance wavelengths $\lambda_0$ of spherical Mie voids within the idealized cavity model via
\begin{equation}
\label{eq:lambda0}
\lambda_0\approx\frac{2\pi n_{\mathrm{in}}R}{y}.
\end{equation}
At finite refractive-index contrast, deviations from this limiting behavior introduce resonance-frequency shifts and radiative losses; thus, Eqs.~\eqref{eq:limit-res}-\eqref{eq:lambda0} provide asymptotic estimates rather than exact resonance wavelengths.\\
The first correction in $m$ gives the radiative loss. Near a root $y_0$ of Eq.~\eqref{eq:limit-res}, we expand the resonance conditions~\eqref{eq:resonance} to first order in $m$, with $\xi_\ell'(x)/\xi_\ell(x)=\iu$. The root moves into the lower half of the complex plane, and
\begin{equation}
\label{eq:Qvoid}
Q\approx G_\ell\,\frac{n_{\mathrm{ex}}}{n_{\mathrm{in}}},\qquad
G_\ell^{\mathrm{TM}}=\frac{y_0^2-\ell(\ell+1)}{2y_0},\qquad
G_\ell^{\mathrm{TE}}=\frac{y_0}{2}.
\end{equation}
The roots of Eq.~\eqref{eq:roots} give $G_1^{\mathrm{TM}}=\text{1.007}$ for the electric dipole (ED) and $G_1^{\mathrm{TE}}= $ 2.247 for the magnetic dipole (MD). For an air void in silicon, $n_{\mathrm{ex}}/n_{\mathrm{in}}=\text{3.5}$, Eq.~\eqref{eq:Qvoid} gives $Q=\text{3.53}$ and 7.86. The exact poles give 3.50 and 7.63 (Fig.~\ref{fig:Fig3}b). 
For TE modes, $G_\ell^{\mathrm{TE}}=y_0/2$ with\linebreak $y_0=n_{\mathrm{in}}k_0R$ is exactly the Fabry-P\'erot estimate of Eq.~\eqref{eq:Q_void}. For TM modes,\linebreak $G_\ell^{\mathrm{TM}}=(y_0/2)\cos^2\theta$ with $\sin\theta=\sqrt{\ell(\ell+1)}/y_0$.  The high-contrast expansion of Hamidi et al.\ gives the same linear scaling.\supercite{Hamidi2025} Its leading term is the normal-incidence limit, $\theta=\text{0}$.
\subsection*{Analogies between Mie resonances and Mie void resonances}
\begin{figure}[t!]
\centering
\includegraphics[width=\linewidth]{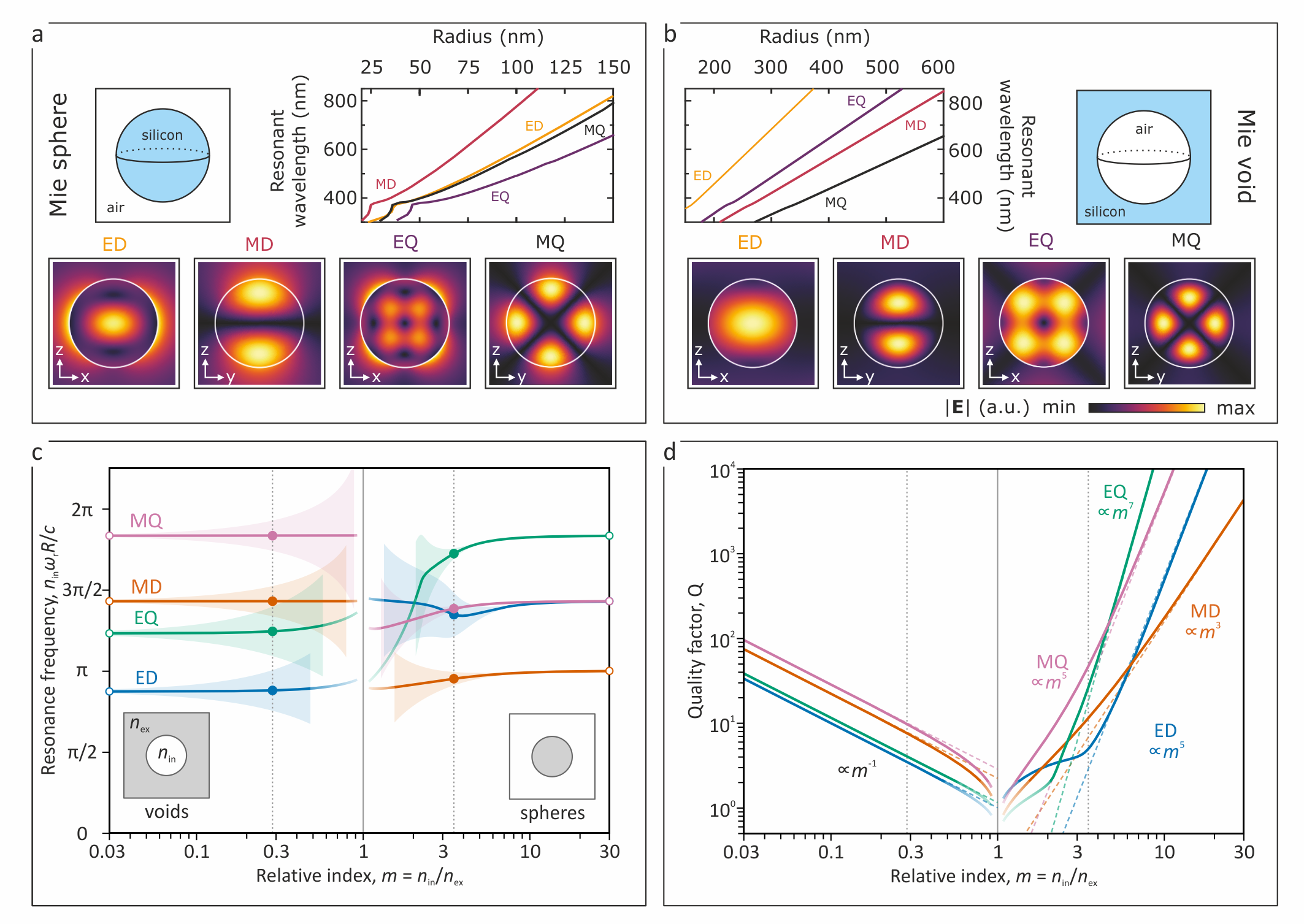}
\caption{\textbf{Comparison of conventional dielectric Mie resonators and dielectric Mie voids.} 
\textbf{a}, Calculated electric and magnetic dipolar and quadrupolar resonances of a silicon Mie sphere in air, showing modal field distributions and dispersion relations. \textbf{b}, corresponding resonances of a spherical air cavity embedded in silicon. Both geometries support localized resonant modes described by Mie theory, despite the inverted refractive-index contrast.
\textbf{c}, Resonance frequency $n_{\mathrm{in}}\omega_{\mathrm{r}}R/c$ of the lowest ED, MD, EQ, and MQ resonances, versus $n_{\mathrm{in}}/n_{\mathrm{ex}}$. Voids are on the left and spheres on the right. Bands: $\omega_{\mathrm{r}}\pm\gamma/2$, from the complex eigenfrequency $\tilde{\omega}=\omega_{\mathrm{r}}-\iu\gamma/2$. Open circles: infinite-contrast limits, Eq.~\eqref{eq:limit-res} for voids. Filled circles: $n_{\mathrm{in}}/n_{\mathrm{ex}}=\text{1/3.5}$ and $\text{3.5}$ (dotted lines), shown in a and b. The colored shaded region indicates the resonance bandwidth. \textbf{d}, Quality factor $Q=\omega_{\mathrm{r}}/\gamma$.  In c and d, lines are dotted and have no band where $Q<\text{2}$. Calculated with MieLib.\supercite{mielib} a, b, Adapted with permission.\supercite{Hentschel2023} Copyright 2023 Springer Nature Limited.}
\label{fig:Fig3}
\end{figure}

The similarities and differences between conventional dielectric Mie resonators and dielectric Mie voids are illustrated in Fig.~\ref{fig:Fig3}.\supercite{Hentschel2023} Using the Mie-theoretical framework of Eqs.~\eqref{eq:a}--\eqref{eq:resonance}, the optical response in terms of scattering cross sections, modal field distributions, and dispersion relations can be calculated. In Figs.~\ref{fig:Fig3}a,b, the modal field distributions for both systems demonstrate that solid and void structures support closely related electric and magnetic dipolar and quadrupolar resonances despite the inverted refractive-index contrast. The Mie sphere modes are localized predominantly inside the silicon particle while simultaneously extending partially into the surrounding air environment. In contrast, the Mie void modes exhibit substantially stronger electric-field confinement within the air cavity itself and only weak modal penetration into the surrounding silicon host. The surrounding dielectric medium therefore primarily defines the optical boundary conditions responsible for resonance formation, whereas the resonant electromagnetic hot spot remains localized inside the accessible low-index cavity.\\
The dispersion relations for solid particles and void systems further emphasize the fundamentally different role of the dielectric environment in both systems.\supercite{Hentschel2023} For silicon Mie spheres, the resonant wavelengths exhibit pronounced bending and spectral restructuring due to the intrinsic dielectric dispersion and absorption resonances of silicon. In particular, the resonances become strongly perturbed near wavelengths around \unit[380]{nm}, where silicon exhibits substantial optical loss. By contrast, the dispersion relations of the Mie void modes remain almost perfectly linear across the investigated spectral range, indicating that the resonant electromagnetic fields are predominantly localized within the air cavity and therefore only weakly affected by the dielectric dispersion of the surrounding host material.\\
At the same time, the inverse dielectric geometry substantially modifies the spectral ordering and spatial confinement of the modes. Whereas the magnetic dipolar mode forms the fundamental eigenmode of the silicon Mie sphere, the electric dipolar mode becomes the lowest-order resonance for the dielectric Mie void. Furthermore, significantly larger cavity dimensions are required for the Mie void resonances to appear within the same spectral range as the corresponding silicon-particle modes. This behavior arises naturally from the large refractive-index difference between silicon and air and the resulting modification of the optical path length inside the resonator. Both lowest-order modes span about one wavelength inside the resonator, $n_{\mathrm{in}}\omega_{\mathrm{r}}R/c\approx\pi$. At the same wavelength, the void ED therefore requires a radius approximately 3.2 times that of the sphere MD.\\
One can investigate these properties in a more general fashion by calculating the resonance frequencies and quality factors as a function of the relative refractive index $n_{\mathrm{in}}/n_{\mathrm{ex}}$, as illustrated in Figs.~\ref{fig:Fig3}c,d. Here, the resonance frequencies are shown as solid lines and the resonance linewidths as shaded areas for the four lowest-order Mie resonances, namely the electric dipole (ED), magnetic dipole (MD), electric quadrupole (EQ), and magnetic quadrupole (MQ). For refractive-index ratios below unity ($n_{\mathrm{in}}/n_{\mathrm{ex}}<\text{1}$), corresponding to low-index inclusions (voids) in a higher-index environment, the resonance frequencies approach approximately constant values as the index contrast increases. In contrast, for conventional high-index spheres ($n_{\mathrm{in}}/n_{\mathrm{ex}}>\text{1}$), the modes exhibit a different ordering and dependence on refractive-index contrast. In particular, the ED and MD modes become nearly degenerate in frequency at large index ratios, whereas the EQ and MQ modes approach different limiting values. The region around $n_{\mathrm{in}}/n_{\mathrm{ex}}=\text{1}$ represents vanishing refractive-index contrast, where the distinction between the inclusion and its surrounding medium disappears, and the resonances cease to be well-defined. Approaching this region, the resonance linewidths increase strongly, indicating a loss of optical confinement.\\
The corresponding quality factors of the resonances, shown in Fig.~\ref{fig:Fig3}d, reveal a pronounced difference between void and sphere resonances. On the low-index side, the quality factors increase approximately as $Q\propto n_{\mathrm{ex}}/n_{\mathrm{in}}$ with increasing contrast. For high-index spheres, the increase is substantially stronger, approaching a cubic dependence, $Q\propto(n_{\mathrm{in}}/n_{\mathrm{ex}})^3$, at sufficiently large index ratios. The quality factors decrease towards values of order unity as the refractive-index contrast vanishes, reflecting the loss of resonant confinement. Thus, although both low-index voids and high-index spheres support multipolar resonances, their confinement properties and scaling with refractive-index contrast differ fundamentally.\\
Comparing the quality factors of spheres and voids reveals the importance of material dispersion and absorption. In silicon Mie spheres, the quality factors decrease strongly toward shorter wavelengths and eventually vanish near the onset of strong intrinsic material absorption in silicon.\supercite{Hentschel2023} The resonant behavior therefore becomes fundamentally limited by the dielectric loss of the high-index material itself. In dielectric Mie voids, however, an entirely different situation emerges. Because the electromagnetic modes remain localized predominantly inside the air cavity, intrinsic absorption in the surrounding silicon host contributes only weakly to resonant damping. Instead, the increase in both the real and imaginary parts of the silicon refractive index near the ultraviolet spectral region enhances Fresnel reflection at the air--silicon interface and thereby strengthens the confinement of the resonant modes inside the cavity. As a result, the Mie void resonances exhibit increasing quality factors toward shorter wavelengths and remain strongly confined even within spectral regions where conventional silicon Mie resonators become heavily damped. The void quality factor grows in proportion to $n_{\mathrm{ex}}/n_{\mathrm{in}}$, Eq.~\eqref{eq:Qvoid}, so a higher host index alone increases it.\\
This behavior highlights one of the central conceptual differences between conventional dielectric resonators and inverse dielectric Mie voids. In dielectric particles, strong field confinement is intrinsically tied to the optical properties and losses of the resonator material itself. In inverse dielectric resonators, by contrast, the surrounding dielectric medium provides confinement while the electromagnetic fields remain concentrated inside a low-index and physically accessible cavity region. The optical properties of the host material therefore influence the resonances primarily through boundary conditions and Fresnel confinement rather than direct modal overlap with absorptive
 material.\\
The observation that the inverse systems voids and sphere exhibit similar resonances is indirect evidence that the "quasi-Babinet" principle has to hold, even though in a strict sense, the exact Babinet principle does not apply. Indeed, Hamidi and co-workers demonstrated that many of the resonant properties of dielectric particles and dielectric voids can be understood as complementary electromagnetic systems.\supercite{Hamidi2025} Unlike the exact Babinet symmetry known from perfect conductors, the dielectric case exhibits only an approximate correspondence, yet the analysis reveals a remarkable connection between the resonant frequencies, quality factors, and modal structures of both geometries. The theory suggests that electric and magnetic resonances exchange their roles upon inversion of the refractive-index contrast, leading to a systematic mapping between resonant modes in particles and voids. Nevertheless, deviations arising from geometrical differences, material dispersion, and absorption limit the validity of this approximation. In particular, reduced loss in the surrounding host material does not necessarily imply loss-free resonant behavior.\\
Fig.~\ref{fig:Fig3}c shows the infinite-contrast form of this mapping. The void MD and the sphere ED reach the same value, $\psi_1(y)=\text{0}$, and so do the void MQ and the sphere EQ, $\psi_2(y)=\text{0}$.  This framework provides an intuitive explanation for the existence of localized resonances in low-index cavities. The interaction between Mie void resonances and dispersive optical materials has also been investigated theoretically, revealing the formation of polaritonic modes and further extending the range of light-matter interactions accessible in inverse dielectric resonators.\supercite{Ryabkov2026}\\
Due to their large size and the lower degree of overall confinement, the quality factors of Mie void resonances are smaller compared to solid Mie resonators. For silicon without loss, $n=\text{3.5}$, the four resonances in Fig.~\ref{fig:Fig3}d have $Q=\text{3.5-9.8}$ for the void and $Q=\text{5.0-47}$ for the sphere. This also leads to a larger spectral width but also to a broader operational range and sizable spectral overlap of the Mie void modes. This behavior results in a broadband generalized Kerker effect, as presented by Reichel and co-workers.\supercite{Reichel2026} In contrast to conventional dielectric Mie resonators, where directional scattering typically emerges only within narrow spectral regions through the overlap of a small number of resonant modes, Mie voids support multiple spectrally overlapping electric and magnetic multipoles. The resulting broadband multipolar interference suppresses backward scattering and produces robust forward-directed radiation. This behavior even persists for local dipole excitation. Depending on the emitter position, Mie voids either enhance spontaneous emission through coupling to air-confined resonant modes or redirect radiation through interference between the dipole and scattered fields. This effect may also contribute to the forward scattering in the textured silicon solar cells as observed by Mann and co-workers.\supercite{Mann2011}
\subsection*{Mie void resonances as cylindrical waveguide modes}
Experimental feasibility requires the fabrication of Mie voids preferentially as cylindrically or conically shaped holes in the substrate surface. In this geometry, an intuitive picture and accurate predictions can be obtained by describing the Mie void as a finite optical waveguide, rather than an isolated scatterer.\supercite{Schwab2026} Within this waveguide-based description, the transverse modal confinement is treated separately from the axial resonance condition. For a cylindrical cavity of radius $a$, depth $d$, and internal refractive index $n_{\mathrm{in}}$, the transverse field distributions are described by Bessel functions of the first kind $J_\ell(\kappa_{\mathrm{in}}r)$,
\begin{equation}
\label{eq:cylinder}
\begin{split}
E_z&\propto J_\ell(\kappa_{\mathrm{in}}r)\,\eu^{\iu\ell\phi}\eu^{\iu k_z z},\\
\kappa_{\mathrm{in}}^2+k_z^2&=n_{\mathrm{in}}^2k_0^2,\\
n_{\mathrm{eff}}&=\frac{k_z}{k_0},
\end{split}
\end{equation}
where $(r,\phi,z)$ are cylindrical coordinates with the $z$-axis along the cavity axis, $\ell$ is the azimuthal order, $\kappa_{\mathrm{in}}$ is the transverse wavenumber inside the cavity, and $k_z$ is the axial propagation constant. Applying the electromagnetic boundary conditions at the cavity sidewalls determines the generally complex effective refractive index $n_{\mathrm{eff}}$. In the low-loss limit, axial resonances approximately satisfy the standing-wave condition
\begin{equation}
\label{eq:standing-wave}
d\Re(k_z)\approx q\pi,
\end{equation}
where $q=\text{1,2,}\ldots$ is the axial order.\\
Owing to the finite refractive-index contrast and the electromagnetic boundary conditions, the supported modes generally exhibit hybrid TE-TM characteristics rather than pure modes. Furthermore, reflection-induced phase shifts and interference with the background field reflected from the substrate influence the resonance positions and spectral line shapes observed experimentally.\\
Under normal incidence, incoming light couples into the waveguide modes and propagates axially in the void. While lateral confinement is achieved through the waveguide modes, axial confinement is achieved through multiple reflections between the upper (air) and lower (substrate) interfaces, analogous to a Fabry-P\'erot cavity. Consequently, the resonant behavior of the void is determined by the waveguide dispersion and the associated effective refractive index of its modes, and the accumulation of propagation phase rather than the excitation of distinct Mie multipoles. In addition to the change in geometry, the shift of the voids to the substrate surface also gives rise to broad background reflections from the substrate surface. As a result, the observed far-field reflectance of cylindrical surface voids is governed by the interference between resonant light leaking from the waveguide modes and the strong off-resonant reflection from the substrate.
\section{Applications of Mie voids}
One of the natural applications of Mie voids (as many other resonators) is to create a lattice such as a metasurface. The reduced effect of material absorption makes Mie voids interesting for metasurface operation in spectral regions that are difficult to access with conventional dielectric resonators. Such metasurfaces are \emph{local} since they are based on the properties of individual resonators. Arslan and co-workers demonstrated reflective Mie void metasurfaces in silicon at a wavelength of \unit[325]{nm}, well within its strongly absorbing ultraviolet regime.\supercite{Arslan2026_2} Elliptical Mie voids were employed as geometric-phase elements, enabling reflective off-axis focusing with numerical apertures (NA) up to 0.2 and diffraction-limited performance up to $\mathrm{NA}=\text{0.15}$. The same phase-control concept was further extended to orbital-angular-momentum generation, demonstrating functional ultraviolet wavefront shaping with Mie void resonators. Complementary work has explored transmission-mode Mie void metasurfaces in silicon-rich nitride, extending their operation to transmissive optical functionalities in the visible spectral range.\supercite{Goldberg2026_02} However, Mie voids may have a wide range of applications solely defined by their unique properties and cavity mechanism for trapping both light and matter.
\subsection*{Sensing and nanoscale detection}
\begin{figure}[t!]
\centering
\includegraphics[width=\linewidth]{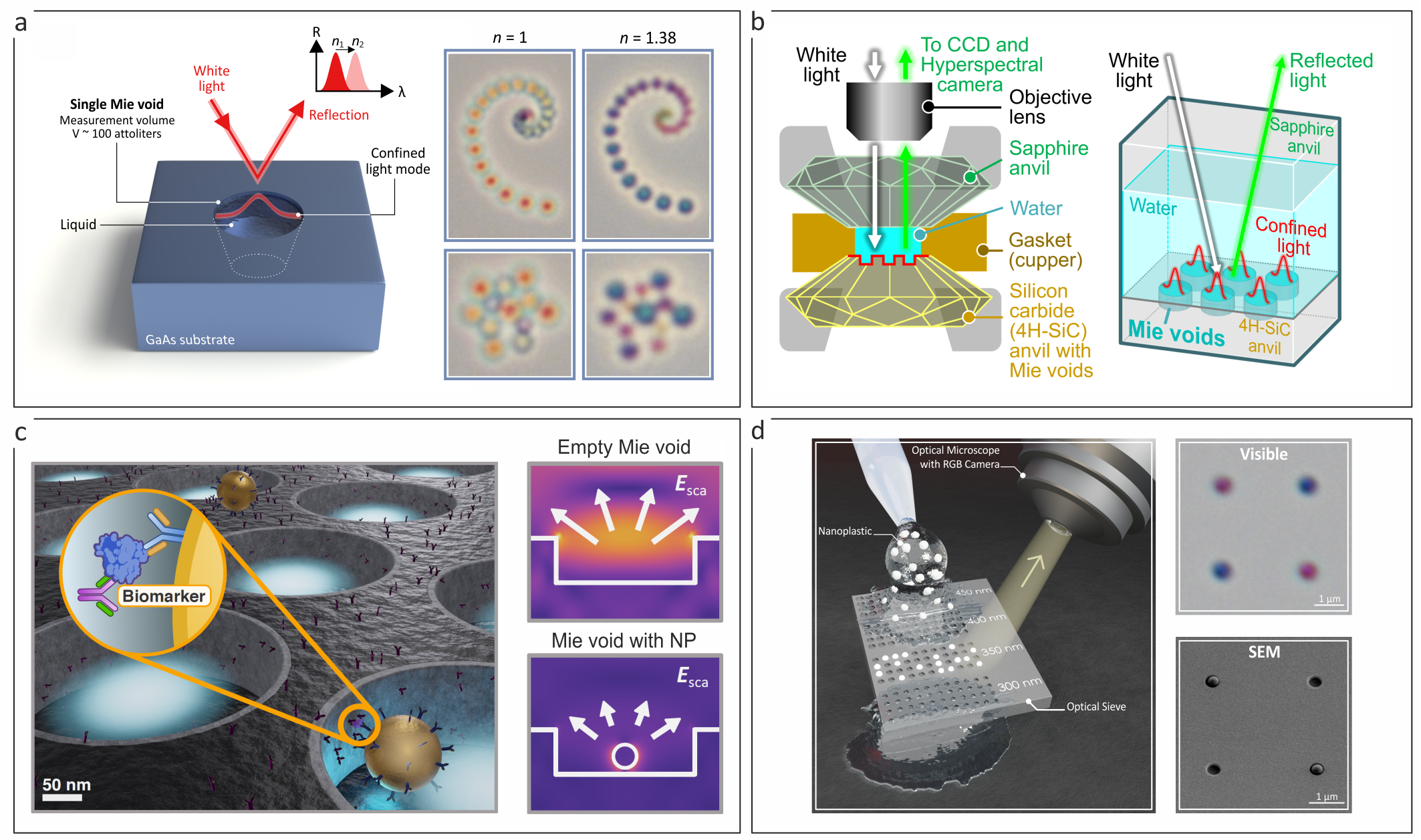}
\caption{\textbf{Sensing applications of Mie voids.} \textbf{a}, Refractive-index sensing using a single Mie void with an attoliter-scale sensing volume. \textbf{b}, High-pressure refractive-index sensing implemented in a silicon-carbide anvil-cell platform. \textbf{c}, Digital biosensing based on the capture and optical detection of functionalized gold nanoparticles within Mie void arrays. \textbf{d}, Optical-sieve concept for nanoplastic detection, sizing, and counting using resonant Mie void arrays. The accessible mode volume of Mie voids enables direct interaction between analytes and the resonant electromagnetic field. a, Adapted with permission.\supercite{Arslan2025} Copyright 2025 American Chemical Society. b, Reprinted with permission.\supercite{Niinomi2025} Copyright 2025 American Chemical Society. d, Adapted with permission.\supercite{Ludescher2025} Copyright 2025 Springer Nature Limited.}
\label{fig:Fig5}
\end{figure}

Sensing has emerged as one of the most extensively explored applications of dielectric Mie void resonators. The platform supports a broad range of approaches, including refractive-index measurements in ultra-small liquid volumes, sensing under extreme conditions, particle detection, and biosensing based on molecular recognition. Despite their different implementations, these concepts rely on the strong perturbation of cavity resonances by changes occurring within the void volume. The concept of Mie void-based sensing has also been extended to the terahertz spectral range, where high-Q resonances have been theoretically investigated for ultrasensitive metasurface sensing applications.\supercite{Kumar2025}\\
The first experimental demonstration of sensing with dielectric Mie voids exploited this direct overlap between the optical mode and the analyte volume for refractive-index sensing in extremely small liquid volumes.\supercite{Arslan2025} By monitoring resonance shifts induced by liquids introduced into the individual cavities, attoliter-scale sensing volumes were realized while maintaining sensitivities comparable to established nanophotonic sensing approaches, see Fig.~\ref{fig:Fig5}a. The Mie void serves a dual function, supporting a cavity mode while simultaneously defining the sensing volume through the void geometry itself.\\
Given that the refractive index is large enough, such a sensor can even be directly integrated into a device. Niinomi and co-workers implemented Mie void resonators in silicon carbide and diamond-anvil-compatible platforms for high-pressure refractive-index sensing of water.\supercite{Niinomi2025} The refractive-index change is monitored under elevated pressures using the spectral response of the void resonances, see Fig.~\ref{fig:Fig5}b. Unlike plasmonic sensing structures, whose optical properties can be modified by pressure-induced changes of particle geometry or electronic response, dielectric void resonators retain their structural integrity and optical functionality under extreme conditions and thus serve as local optical probes in environments where conventional nanophotonic sensing approaches become increasingly difficult to  implement.\\
For many advanced sensing applications, it is not sufficient to simply detect refractive index shifts, but functionalization is required. Riabov and co-workers introduced functionalized gold nanoparticles into arrays of Mie voids, see Fig.~\ref{fig:Fig5}c.\supercite{Riabov2026} The presence of the particle modifies the optical response and scattering significantly, and thus the overall signal becomes highly sensitive to binding events on the gold particles themselves. Rather than monitoring ensemble-averaged resonance shifts, this approach detects and counts individual nanoparticle binding events associated with molecular recognition processes utilizing image-analysis algorithms to identify individual binding events across large sensing areas. As a proof of concept, the platform was used for the detection of the inflammatory biomarker interleukin-6 at physiologically relevant concentrations.\\
Beyond refractive-index measurements, Mie voids have also enabled sensing concepts based on the interaction between resonant cavities and individual particles.\supercite{Ludescher2025} An example is provided by the optical sieve architecture, in which arrays of dielectric void resonators act simultaneously as particle filters and optical reporters. Nanoplastic particles trapped within appropriately sized cavities perturb the cavity resonance and generate distinct color changes that can be recorded using conventional optical microscopy. Because particle occupancy depends on cavity dimensions, while the optical response depends on the dielectric perturbation introduced by the trapped particle, the platform provides information on particle size, particle number, and concentration within a single measurement. This concept can even work with environmentally relevant samples consisting of heterogeneous materials and particle shapes, pushing it towards all-optical nanoplastic detection.\supercite{Ludescher2025, Krauss2025, Ludescher2026} Further improvements in the detection efficiency of the optical sieve have been explored through the application of principal component analysis to the optical response of the Mie void arrays.\supercite{Ludescher2026_02}\\
Taken together, these studies illustrate how sensing with dielectric Mie voids has progressed far beyond simple refractive-index measurements. The same resonant platform has been adapted to operate under extreme conditions, detect individual nanoparticles, and transduce molecular recognition events into optical signals. This evolution reflects a broader shift from measuring bulk material properties toward probing increasingly localized and discrete interactions within the cavity volume.
\subsection*{Active and quantum photonics}
\begin{figure}[t!]
\centering
\includegraphics[width=\linewidth]{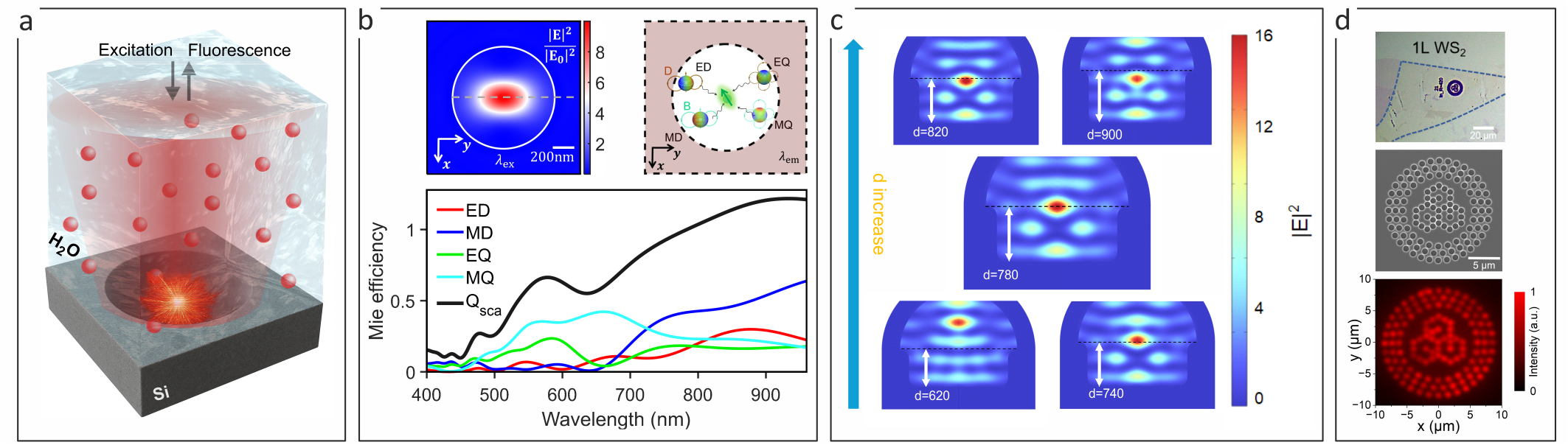}
\caption{\textbf{Active and quantum photonics with Mie voids.} \textbf{a}, Concept of fluorescence enhancement using emitters located within the resonant volume of a Mie void. \textbf{b}, Presentation of excitation and emission enhancement mechanisms in dielectric Mie voids. \textbf{c}, Simulated electric-field distributions illustrating the spatial overlap between Mie void resonances and monolayer WS\textsubscript{2} emitters. \textbf{d}, Demonstration of a nanoscale excitonic display based on arrays of Mie void resonators integrated with two-dimensional semiconductors. These examples highlight the use of Mie voids for fluorescence enhancement, spontaneous-emission control, and excitonic light sources. a, Reprinted with permission.\supercite{Kuznetsov2025} Copyright 2025 MDPI. d, Adapted with permission.\supercite{Liao2026} Copyright 2026 American Chemical Society.}
\label{fig:Fig6}
\end{figure}

Controlling the interaction between optical resonances and quantum emitters is a central objective of modern nanophotonics. By modifying the local electromagnetic environment, resonant nanostructures can influence excitation rates, spontaneous-emission dynamics, emission patterns, and nonlinear optical processes. Mie void resonators provide a particularly interesting platform for exploring these effects because emitters can be positioned directly within regions of strong optical fields.\\
An early demonstration of these capabilities was provided through fluorescence spectroscopy using silicon Mie voids, see the artistic illustration in Fig.~\ref{fig:Fig6}a.\supercite{Kuznetsov2025} As can be seen from the simulated field distributions, the structures are uniquely suited to interact with the single fluorophores, biomolecular conjugates, and quantum dots utilized in this study. The authors were able to combine resonant excitation enhancement with moderate Purcell enhancement and improved photon collection, while maintaining wavelength-scale detection volumes.\\
A more general framework for emitter-Mie void interactions was later developed by Fu and co-workers through the concept of quantum-emission engineering with Mie voids.\supercite{Fu2026_02, Fu2026} In this work, fluorescence enhancement was decomposed into separate excitation and quantum-yield contributions, allowing the optical response of emitter-cavity systems to be predicted quantitatively. The analysis confirms that the overall enhancement can be expressed as the product of excitation enhancement and quantum-yield enhancement, thereby separating the effects of local-field enhancement from modifications of spontaneous-emission dynamics. Systematic measurements on large parameter sweeps of cavity geometries confirmed the predictive power of this framework and demonstrated significant reductions in fluorescence lifetimes consistent with Purcell-enhanced spontaneous emission. Importantly, the authors also found significant differences in the linewidth of the modes and thus a \enquote{broadband} interaction of the modes in Mie voids (see Fig.~\ref{fig:Fig6}b), compared to the case of solid Mie  resonators.\\
The direct accessibility of the optical mode volume becomes even more advantageous when coupling resonances to atomically thin quantum materials.\supercite{Lu2026,Sarbajna2025,Liao2026} Monolayer transition-metal dichalcogenides can be positioned directly at the dielectric-air interface where the optical field reaches its maximum. Additionally, this geometry avoids stress and strain in the material, which are known to modify the optical response considerably. The simulations displayed in Fig.~\ref{fig:Fig6}c demonstrate that by precisely tuning the depth of the voids, the field maximum can be shifted over the interface, maximizing interaction with the van der Waals material. Lu and co-workers demonstrated that combining monolayer WS\textsubscript{2} with Mie voids allows for resonant coupling between the cavity modes and excitonic transitions, resulting in approximately twenty-fold photoluminescence enhancement and more than an order-of-magnitude enhancement of second-harmonic generation. The spatial localization of the optical modes furthermore enabled direct imaging of the cavity field distribution through nonlinear optical microscopy, providing a direct visualization of the emitter-cavity interaction. Sarbajna and co-workers studied a very similar system consisting of Mie voids directly made of bulk WSe\textsubscript{2} with a monolayer of WS\textsubscript{2} on top.\supercite{Sarbajna2025} In reference to the WSe\textsubscript{2} monolayer on the material itself, they observe fluorescence enhancement of about 600. Interestingly, in taking back-focal-plane images, the authors can also experimentally demonstrate the modification of the photoluminescence emission pattern, demonstrating efficient coupling to the nanophotonic mode. Liao and co-workers studied the coupling of monolayer WS\textsubscript{2} with Mie voids in silicon (Fig.~\ref{fig:Fig6}d).\supercite{Liao2026} The authors point out that the interface-induced exciton suppression is a major hurdle in many nanophotonic coupling schemes. Utilizing Mie voids eliminates this issue altogether due to the free-standing monolayer material on top of the voids. In this experiment, enhancement factors on the order of 120 are observed. Comparing these three experiments, one has to notice the different measured enhancement factors. In fact, such enhancements are very difficult to compare among experiments as they depend on the utilized collection NA, the normalization to \enquote{bare material on substrate} or to \enquote{off-resonant free-standing}, and more.\\
Taken together, these studies illustrate the flexibility with which emitter-cavity interactions can be tailored in inverse dielectric resonators. Depending on the emitter system and cavity design, the resulting optical response can be dominated by excitation enhancement, modified spontaneous emission, nonlinear optical effects, or changes in emission directionality. Complementary numerical studies have also investigated the optical Kerr nonlinearity in high-index metasurfaces containing Mie void arrays, exploring the role of resonant field distributions and inter-void coupling in the effective nonlinear optical response.\supercite{Panov2026} The diversity of observed phenomena underscores the broad parameter space available for engineering light-matter interactions in these systems.
\section{Fabrication of Mie voids}
\begin{figure}[t!]
\centering
\includegraphics[width=\linewidth]{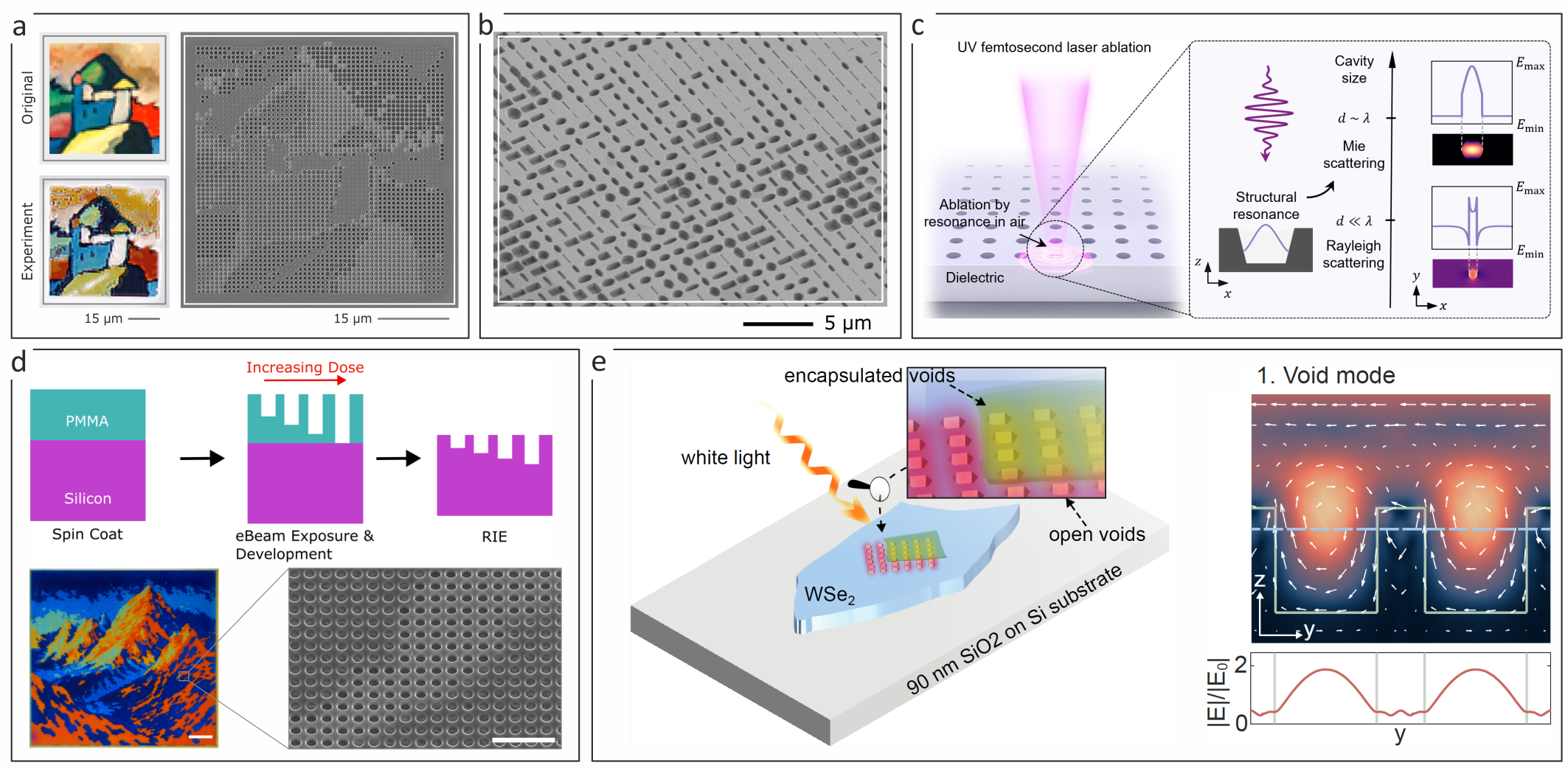}
\caption{\textbf{Fabrication approaches for Mie voids.} \textbf{a}, Structural-color printing based on focused-ion-beam fabricated Mie voids with varying diameters and depths. \textbf{b}, Multilayer electron-beam lithography and reactive-ion etching enabling resonator arrays with multiple discrete cavity depths. \textbf{c}, Femtosecond laser ablation is employed to fabricate Mie void resonators in silicon. \textbf{d}, Single-step grayscale electron-beam lithography for the fabrication of continuously varying depth profiles and Mie void metasurfaces. \textbf{e}, Mie void resonators fabricated in van der Waals materials and subsequently modified through heterostructure assembly and encapsulation. The different approaches provide complementary capabilities in terms of geometric control, scalability, and material integration. a, Reprinted with permission.\supercite{Hentschel2023} Copyright 2023 Springer Nature Limited. b, Reprinted with permission.\supercite{Arslan2026} Copyright 2026 John Wiley \& Sons, Inc. d, Adapted with permission.\supercite{Goldberg2026} Copyright 2026 American Chemical Society. e, Adapted with permission.\supercite{Sarbajna2025_2} Copyright 2025 John Wiley \& Sons, Inc.}
\label{fig:Fig7}
\end{figure}

Mie void resonators offer two distinct spectral tuning parameters, which are depth and diameter. As they are fabricated into a bulk substrate, both parameters can be independently tuned without sacrificing integrability or other practical design constraints. This, however, does require slightly modified fabrication compared to other nanophotonic systems.\supercite{Yang2025}\\
The first experimental realizations of Mie voids employed focused ion beam (FIB) milling of cavities into bulk silicon.\supercite{Hentschel2023, Graydon2023} FIB fabrication provides direct control over cavity diameter and depth through the local ion dose and is thus the ideal prototyping technique.\supercite{Kuznetsov2025} Fig.~\ref{fig:Fig7}a depicts the use of Mie voids for nanoscale color printing. To realize a broad range of colors and saturation, one needs to vary the diameter and depth, as can be seen in the scanning electron micrographs. The very fine depth variation can straightforwardly be implemented with FIB milling.\supercite{Fu2026,Radhakrishnan2025} Compared to reactive ion etching of semiconductor structures, one does, however, observe truncated conical profiles which modify the optical response slightly compared to the perfect rectangular shape assumed in most simulations. One key benefit of FIB is the ability to structure literally every material of interest with little optimization effort. This ability also becomes highly relevant for unusual size and shape substrates, as in the case of the 4H-SiC anvil cells used for high-pressure Mie void sensing.\supercite{Niinomi2025}\\
One significant drawback is the serial nature of FIB fabrication, which severely limits throughput and large-area fabrication. A straightforward improvement is the use of electron-beam lithography with reactive dry etching methods, which are well established in semiconductor processing and nanophotonics.\supercite{Yang2025} Generally, this results in one fixed depth for all structures, thus restricting the spectral tuning to lateral size variation.\supercite{Maruyama2025} For many applications in sensing or metasurface design, this does not pose significant limitations, but can, however, be improved by the use of multiple subsequent electron-beam lithographic and dry etching steps. An example is depicted in Fig.~\ref{fig:Fig7}b. Here the authors realized four distinct etching and structure depths by a four-layer  process.\supercite{Arslan2026}\\
Recently, a novel fabrication concept was demonstrated that directly uses Mie void resonance during structure formation (Fig.~\ref{fig:Fig7}c). Gong and co-workers proposed a femtosecond-laser-based approach termed Mie-lithography, in which an initial laser-ablated nanocavity acts as a seed for the subsequent fabrication process.\supercite{Gong2026} Once the cavity reaches dimensions supporting a Mie resonance, the optical field becomes strongly localized inside the void and drives further material removal, establishing a feedback process between the evolving cavity geometry and the resonant mode. The authors demonstrated that ultraviolet excitation at \unit[343]{nm} can sustain this resonance-assisted vertical growth. In contrast, non-resonant excitation at \unit[800]{nm} mainly leads to lateral expansion despite the larger optical penetration depth of silicon at this wavelength. The resulting geometry can be controlled through the laser parameters, with pulse number mainly determining the cavity depth, pulse energy tuning the lateral dimensions, and polarization controlling the in-plane symmetry and orientation. This single-step approach therefore enables multidimensional control of Mie void geometries without subsequent lithography and etching steps.\\
While these methods are in general scalable, the depth variation is limited. Continuous depth control, however, is desirable as it offers an independent tuning knob. Goldberg and co-workers utilized grayscale lithography to achieve this goal, as illustrated in Fig.~\ref{fig:Fig7}d.\supercite{Goldberg2026} Here, the local exposure dose determines the remaining resist thickness after development, which is subsequently transferred into the substrate during etching. This process enables continuous variation of the cavity depths within a single fabrication cycle. Because the resonance wavelength of Mie voids depends strongly on cavity depth, grayscale lithography provides a direct way to continuously vary optical responses, including depth-gradient metasurfaces and structural-color patterns.\\
Another very interesting fabrication route is related to van der Waals materials (Fig.~\ref{fig:Fig7}e).\supercite{Lu2026,Sarbajna2025,Sarbajna2025_2} This class offers a wide range of high-index materials ideally suited for Mie void fabrication. In principle, the structuring of the voids inside the material can be done with FIB milling or dry etching. One unique feature is the straightforward integration of monolayer van der Waals materials. Conceptually, it is a way to realize a closed Mie void cavity by sealing the voids with stamp-transferred layers of van der Waals materials. Moreover, it is an ideal platform for the integration of emitters found in this material class. Sarbajna and co-workers also addressed another fabrication issue in this material platform which had not been considered before.\supercite{Sarbajna2025} The exact etching depth is in fact not independent of the lateral size: larger structures will also become deeper (modified in- and outflow of etching gases and removed material). The authors studied this behavior in detail, enabling precise depth control irrespective of lateral size.
\section{Summary and outlook}
The examples discussed throughout this review illustrate the diverse ways in which optical resonances can emerge within low-index regions embedded in higher-index environments. While hollow-core waveguides, photonic bandgap structures, antiresonant architectures, inverse dielectric resonators, and Mie voids are governed by distinct physical mechanisms and originated in largely separate research communities, their comparison reveals a recurring theme: optical confinement and resonant behavior can be engineered through the surrounding medium as much as through the material itself.\\
Within this broader context, Mie voids provide a particularly versatile realization of inverse dielectric resonances. Their development has progressed from fundamental studies of optical modes in spherical cavities to applications in sensing, particle characterization, light-matter interaction, nonlinear optics, and quantum photonics. Together, these works demonstrate that inverse dielectric resonators are not merely complementary counterparts to conventional dielectric nanostructures, but constitute a rich class of photonic systems with their own design opportunities and functionality. Importantly, this underpins that Mie void resonators are not meant to replace their solid counterparts but to amend the toolkit available for advanced nanophotonic devices by their complementary properties.\\
The rapid development of fabrication strategies has further expanded the scope of these structures. Grayscale lithography, multilayer fabrication approaches, and heterostructure assembly increasingly allow resonator geometries to be engineered with high precision and integrated into larger photonic architectures. At the same time, the extension of Mie voids beyond silicon into wide-bandgap semiconductors, emerging quantum materials, and heterogeneous material platforms suggests that the underlying concepts are broadly applicable rather than tied to a specific technological implementation.\\
Looking forward, some of the most interesting opportunities may arise from revisiting concepts that have become central to modern nanophotonics within inverse dielectric geometries. Collective lattice resonances, bound states in the continuum, topological photonics, non-Hermitian phenomena, strong-coupling physics, active metasurfaces, and quantum-photonic architectures have only begun to be explored in inverse dielectric resonators. Whether these effects simply translate to complementary structures or give rise to qualitatively new behavior remains largely unknown.\\
More generally, Mie voids provide a useful lens through which to view a broader class of photonic systems in which resonant behavior is shaped by the interplay between material and void, inclusion, and host. Exploring this perspective may reveal new opportunities for controlling optical fields and light-matter interactions, while also providing deeper insight into how electromagnetic states emerge in structured dielectric environments.
\subsection*{Acknowledgments}
\textbf{Funding:} M.H. acknowledges a grant through the Ministerium für Wissenschaft, Forschung und Kunst Baden-Württemberg (RiSC), Vector Stiftung (MINT Innovations), and Carl-Zeiss-Stiftung (CZS Center QPhoton). Y.K. acknowledges support from the Australian Research Council (Grant DP210101292) and the International Technology Center Indo-Pacific (ITC IPAC) via Army Research Office (contract FA520923C0023).\\ \\
\textbf{Author contributions:} All authors contributed to the conception, preparation, writing, and revision of the review manuscript.\\ \\
\textbf{Competing interests:} The authors declare that they have no competing interests. \\
\\
\textbf{Data, code, and materials availability:} All data and code needed to evaluate the conclusions are present in the paper. This study did not generate new materials. No custom code was generated for this study.
\printbibliography
\end{document}